\documentstyle{elsart}

\input epsf

\begin{document}

\begin{frontmatter}

\title{
First Results on the \\
Performance of the HEGRA IACT Array}

\author[3]{A. Daum},
\author[3]{G. Hermann},
\author[3]{M. Hess},
\author[3]{W. Hofmann},
\author[3]{H. Lampeitl},
\author[3]{G. P\"uhlhofer},
\author[3]{F. Aharonian},
\author[4]{A.G. Akhperjanian},
\author[1,5]{J.A. Barrio},
\author[4]{A.S. Beglarian},
\author[3]{K. Bernl\"ohr},
\author[5]{J.J.G. Beteta},
\author[1]{S.M. Bradbury},
\author[5]{J.L. Contreras},
\author[5]{J. Cortina},
\author[2]{T. Deckers},
\author[1]{E. Feigl},
\author[1,5]{J. Fernandez},
\author[5]{V. Fonseca},
\author[3]{A. Fra\ss},
\author[6]{B. Funk},
\author[5]{J.C. Gonzalez},
\author[7]{G. Heinzelmann},
\author[3]{M. Hemberger},
\author[3]{A. Heusler},
\author[1]{I. Holl},
\author[7]{D. Horns},
\author[3,4]{R. Kankanyan},
\author[2]{O. Kirstein},
\author[3]{C. K\"ohler},
\author[3]{A. Konopelko},
\author[1]{D. Kranich},
\author[7]{H. Krawczynski},
\author[1]{H. Kornmayer},
\author[7]{A. Lindner},
\author[1]{E. Lorenz},
\author[6]{N. Magnussen},
\author[6]{H. Meyer},
\author[1,5,4]{R. Mirzoyan},
\author[6]{H. M\"oller},
\author[5]{A. Moralejo},
\author[5]{L. Padilla},
\author[3]{M. Panter},
\author[1]{D. Petry},
\author[1]{R. Plaga},
\author[7]{J. Prahl},
\author[1]{C. Prosch},
\author[2]{G. Rauterberg},
\author[6]{W. Rhode},
\author[7]{A. R\"ohring},
\author[4]{V. Sahakian},
\author[2]{M. Samorski},
\author[5]{J.A. Sanchez},
\author[7]{D. Schmele},
\author[2]{W. Stamm},
\author[3]{M. Ulrich},
\author[3]{H.J. V\"olk},
\author[6]{S. Westerhoff},
\author[6]{B. Wiebel-Sooth},
\author[3]{C.A. Wiedner},
\author[2]{M. Willmer},
\author[3]{H. Wirth}

\collab{HEGRA Collaboration}

\address[3]{Max-Planck-Institut f\"ur Kernphysik, P.O. Box 103980,
        D-69029 Heidelberg, Germany}
\address[4]{Yerevan Physics Institute, Yerevan, Armenia}
\address[1]{Max-Planck-Institut f\"ur Physik, F\"ohringer Ring 6,
        D-80805 M\"unchen, Germany}
\address[5]{Facultad de Ciencias Fisicas, Universidad Complutense,
         E-28040 Madrid, Spain}
\address[2]{Universit\"at Kiel, Inst. f\"ur Kernphysik,
       Olshausenstr.40, D-24118 Kiel, Germany}
\address[6]{BUGH Wuppertal, Fachbereich Physik, Gau\ss str.20,
        D-42119 Wuppertal, Germany}
\address[7]{Universit\"at Hamburg, II. Inst. f\"ur Experimentalphysik,
       Luruper Chaussee 149, D-22761 Hamburg, Germany}

\begin{abstract} First results concerning the performance
characteristics of the HEGRA IACT array are given, based 
on stereoscopic observations of the Crab Nebula with four
telescopes. The system provides a $\gamma$-ray
energy threshold around 0.5~TeV.
The Crab signal demonstrates an angular resolution
of about $0.1^\circ$. Shape cuts allow to suppress cosmic
ray background by almost a factor 100, while maintaining
40\% efficiency for $\gamma$-rays. 
The Crab signal is essentially background
free. For longer observation times of order 100~h, the system
in its present form provides sensitivity to point sources
at a level of 3\% of the Crab flux. Performance is expected to
improve further with the inclusion of the fifth telescope
and the implementation of advanced algorithms for shower
reconstruction.

\end{abstract}

\end{frontmatter}

Imaging Atmospheric Cherenkov Telescopes
(IACTs) have proven a very powerful tool for the 
investigation of galactic and extragalactic
VHE $\gamma$-radiation, see e.g.~\cite{weekes}.
Further advances in the IACT technique are 
expected in two directions: the use of very large
reflector areas to significantly lower the
energy threshold, and the coincident operation
of several IACTs to improve the reconstruction
of shower parameters and, for larger arrays of
IACTs, to significantly increase the effective 
detection area for $\gamma$-rays.

The HEGRA system of IACTs~\cite{hegra_system}
emphasizes the stereoscopic
observation of air showers with an array of telescopes, as
opposed to single or independent IACTs. Stereoscopic 
observation with an array of IACTs promises advantages in different
areas \cite{hegra_system,system1,system2}:
\begin{itemize}
\item Lower trigger thresholds: a trigger coincidence between
multiple distributed telescopes should drastically reduce
the number of random coincidences caused by background light, 
and the rate of triggers by local muons. As a result, the 
individual telescopes can be operated with reduced pixel 
trigger tresholds, and hence with reduced energy thresholds.
\item Unambiguous spatial reconstruction of the shower axis
and improved angular resolution: the simultaneous
observation of air showers by at least two telescopes is 
expected to measure the direction of a shower with a 
resolution of about $0.1^\circ$, and the core location within
10 to 15~m. \footnote{For single IACTs, source locations
were so far not derived on an event-by-event basis, but 
rather by averaging over event samples with suitable
techniques (see e.g. \cite{whipple_source}). Recently,
methods were discussed which allow to determine directions for
individual showers, with certain limitations (e.g., \cite{cat,ulrichphd}).}
\item Improved energy resolution: the known distance to the
reconstructed shower core should allow to efficiently relate 
light yields to energies, with an energy resolution around
20\% to 25\%. Simultaneous observation with multiple telescopes
makes it likely that for each shower at least one telescope is
in the optimum distance range, where fluctuations are minimal.
\item Improved hadron rejection: while shower shapes in different
views are not completely uncorrelated, one expects a further
increase in hadron rejection based on shape cuts in multiple
views.
\end{itemize}
The HEGRA IACT array allows for the first time to test these
concepts under realistic conditions. 
The Crab Nebula, which was established by the
Whipple group as a VHE $\gamma$-ray source
\cite{whipple_crab} and has since detected and
studied by many groups operating IACTs
(see e.g. \cite{crab1,crab2,crab3,crab4,crab5,crab6,crab7}) provides
a ``standard candle'' in the VHE sky, and is used to
evaluate the performance of the array. 
The HEGRA IACT array is supposed
to ultimately provide sensitivity to sources of about
1/10 of the Crab's strength, corresponding to a flux around
$10^{-12}$/cm$^2$s above 1~TeV~\cite{hegra_system}. 
Even though the array is not
quite in its final shape, and is just starting regular operation,
first results are very encouraging and demonstrate the
power of the stereoscopic approach.

The paper is organized as follows: in the next section,
the HEGRA IACT array is described briefly. Then, 
properties of the multi-telescope coincidence trigger and
the resulting trigger thresholds are addressed. Analysis
procedures used for the Crab data are summarized, and finally
the performance of the IACT array in terms of angular
resolution, background suppression, and minimal detectable
flux is discussed.

\section{The HEGRA IACT Array}

The HEGRA IACT system is located on the Canary Island of La Palma, 
at the Observatorio del Roque de los Muchachos
of the Instituto Astrofisico de Canarias,
at a height of about 2200~m asl.
In its final form, the HEGRA IACT array will consist of 
five identical telescopes with 8.5~m$^2$ mirror area, 
5~m focal length, and 271-pixel cameras with a pixel
size of $0.25^\circ$ and a field of view of $4.3^\circ$
\footnote{The actual field of view is hexagonal; given 
here is
the diameter of a circle with the equivalent area.}.
Four telescopes are arranged in the corners of a square with 
roughly
100~m side length, the fifth telescope is located 
in the center of the square.
The cameras are read out by Flash-ADCs. The two-level
trigger  requires a coincidence of two neighboring pixels
to trigger a telescope, and a coincidence of at least two
telescope triggers to initiate the readout.

\begin{figure}[tb]
\begin{center}
\mbox{
\epsfxsize9.0cm
\epsffile{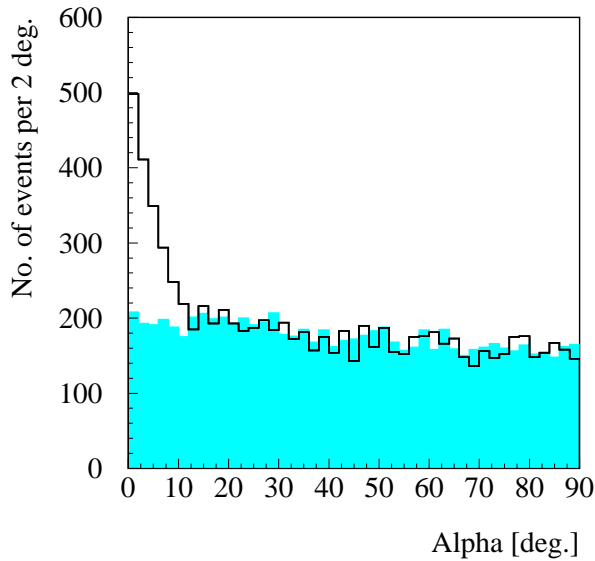}}
\end{center}
\caption
{\em Crab signal observed in 40 h of on-source data taken 
with the telescope CT3 in stand-alone mode, after cuts on
the image parameters {\em width}, {\em length},
{\em concentration}, and {\em distance}. Shown is as
a full line the
distribution in the angle $\alpha$ between the major axis
of the Cherenkov image and the direction to the image of the
source in the center of the camera. The shaded area indicates
the background obtained from an equivalent amount of off-source
observations.}
\label{ct3_crab}
\end{figure}

The first telescope with the final 271-pixel camera
(CT3)
came into operation late in 1995, and clear signals
for $\gamma$-ray emission from the Crab Nebula were
seen in the 1995 and 1996 single-telescope 
data, see Fig.~\ref{ct3_crab}.
The data discussed in the following were taken in Dec. '96
and Jan. '97, shortly after the commissioning of the
next three telescopes CT4,5,6 and their electronics. During this period,
four of the five telescopes
(CT3,4,5,6)
were operational with the 271-pixel cameras. Three of
these used the final 120~Mhz VME Flash-ADC readout; one 
(CT4) was 
still equipped with an interim 40~MHz VME Flash-ADC system.
The lower sampling frequency results in increased
integration times and somewhat higher sky noise for this
telescope. The data presented here were taken with a simple
two-pixel concidence for the telescope trigger, without
a next-neighbor requirement. 
The fifth system telescope (CT2) was still equipped 
with an old, coarser 61-pixel camera and CAMAC ADCs. It was
operated independently from the other telescopes and is 
not used in the analysis discussed in the following. The same is true for
the smaller prototype telescope CT1.
Use of four rather than five telescopes
reduces the rate by 20\% and 30\% for two-telescope
and three-telescope triggers, respectively.
In addition, for the data set used in the analysis
presented here, the central telescope (CT3) was always
required among the triggering telescopes, resulting
in an additional rate loss of about 
25\% compared to the normal situation,
where any two telescopes would trigger the IACT array.

Given the short time since the data was taken, and the
emphasis on continued installation and tests of the
hardware, the software algorithms for data analysis
as well as our understanding of the properties of the
system and its optimization are still imperfect. 
Also, geometric calibration data were not yet
available in sufficient quantity and quality for all
telescopes.

Given these various caveats, the following performance
figures, though quite remarkable, 
should not be taken to reflect the ultimate 
performance of such an
array; one should expect detection rates to increase
by a factor 1.5 to 2, with a corresponding increase
in sensitivity.

\section{Triggering of the HEGRA IACT Array}

The individual HEGRA telescopes are triggered if
two pixels exceed a preset threshold, within a
coincidence window of about 12~ns. Presently,
any two pixels may trigger the
telescope. In future, a hardware next-neighbor
unit will require that at least two trigger pixels
are direct neighbors, to further reduce the rate
of random coincidences. In earlier single-telescope
observations, the next-neighbor check
was emulated in software, providing a 
trigger verification after about 100~$\mu$s. This
delay is however not compatible with the timing
of the multi-telescope coincidence.

The individual telescope trigger signals are routed to
a central station, where they are delayed
in a custom designed VME unit, to compensate differences
in timing which arise since the Cherenkov light front
does not reach all telescopes at the same time. The
delay values are updated continuously as the 
source moves across the sky.
To trigger
the readout of the IACT array, at least two
telescopes have to trigger within a time
window of 70~ns; this window is large compared
to the timing jitter between telescope signals,
of order 10~ns. An inter-telescope coincidence
causes a global trigger signal and an event number to
be transmitted back to all telescopes, including
those which did not trigger themselves. After
such a global trigger, the Flash-ADCs stop
recording signals, the local VME CPU at each
telescope locates
the appropriate time slice in the Flash-ADC memory,
and reads out the data, with on-line sparsification.
Data are buffered and multi-event packets
are sent via Ethernet to a central event builder
station.

\begin{figure}[tb]
\begin{center}
\mbox{
\epsfxsize10.0cm
\epsffile{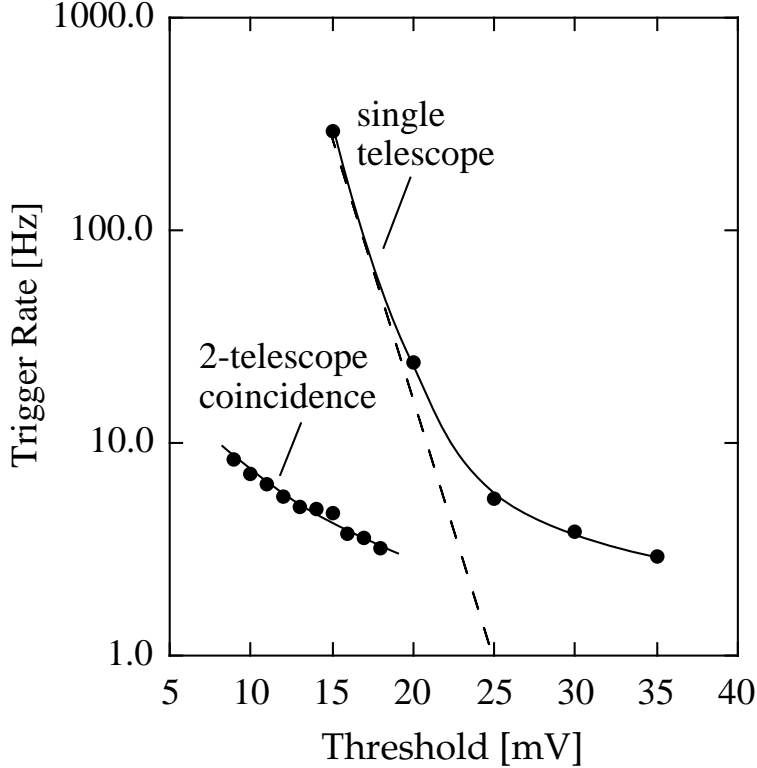}}
\end{center}
\caption
{\em Rate of single-telescope (CT3) triggers and of 
two-telescope (CT3,4) coincidences, as a function of 
the pixel trigger threshold applied to the camera 
pixels. One photoelectron is roughly equivalent
to 1.2~mV. The dashed line shows the calculated rate of
random single-telescope triggers. The 
rate of random two-telescope coincidences is small
compared to the measured rates.}
\label{fig_rates}
\end{figure}

Fig.~\ref{fig_rates} shows the trigger rates for
one single telescope (CT3) and for a 
coincidence of two given telescopes (CT3,4), as a function
of the pixel trigger threshold in mV. One photoelectron
corresponds roughly to 1.2~mV. For low tresholds,
the single-telescope rate shows a very steep 
dependence on the pixel trigger threshold. In
this regime, triggers are almost entirely caused 
by random coincidences due to night-sky photons;
the observed rate is fully consistent with the
expected random rate calculated from the measured
single-pixel rates and the width of the coincidence
window.
Only for thresholds above 25~mV the dependence
flattens out, and genuine air-shower triggers 
dominate. In contrast, the two-telescope trigger rate
is over the entire range, down to pixel thresholds of
8~mV, governed by air showers, and exihibits a
slope similar to the integral cosmic-ray
spectrum. For such a trigger, the choice of the
pixel thresholds is no longer determined purely by
the onset of noise triggers, but rather by other
considerations such as the minimal light yield
required to generate a useable image.

For any given pixel threshold, the
rate of two-telescope (CT3,4) coincidences is smaller than
the rate of good single-telescope (CT3) triggers. The reason
is twofold: on the one hand the effective area of
a two-telescope coincidence is smaller than that of
a single telescope, since both telescopes have
to be located within the light pool of an air shower.
On the other hand, the coincidence requirement 
provides already at the trigger level 
a suppression of hadron-induced
showers as compared to $\gamma$-ray showers.

\begin{figure}[tb]
\begin{center}
\mbox{
\epsfxsize10.0cm
\epsffile{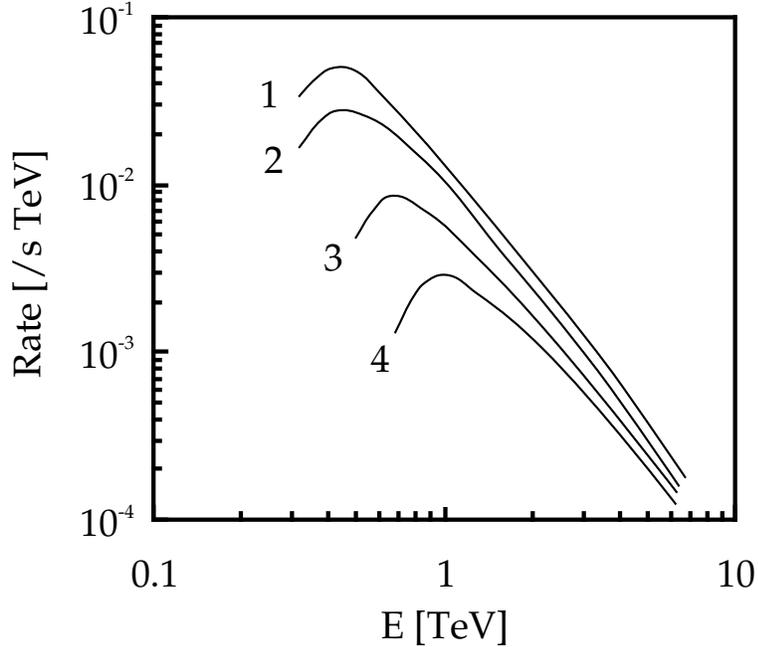}}
\end{center}
\caption
{\em 
Simulated differential detection rates for a $\gamma$-ray source
with a differential spectral index of 2.7 and a flux of
$10^{-11}$/cm$^2$s. Curves: (1) full IACT system, 8-photoelectron
threshold, (2) IACT system as used in the present analysis,
8-p.e. threshold, 
(3) single telescope, 15-p.e. threshold, (4) single telescope,
22-p.e. threshold.}
\label{fig_drates}
\end{figure}

\begin{table} [tb]
\begin{center}
\begin{tabular}{|l|c|c|}
\hline
Mode & Pixel threshold & Energy threshold \\
  & (photoelectrons) & (TeV) \\
\hline \hline
Single telescope, & $\approx 22$ & $\approx 1.0$\\
~~~any 2 pixels & & \\ \hline
Single telescope, & $\approx 15$ & $\approx 0.7$ \\
~~~two neighbor pixels & & \\ \hline
4-Telescope system, & $\approx 8$ & $\approx 0.5$ \\
~~~at least two telescopes, & & \\
~~~any two pixels per telescope & & \\
\hline
\end{tabular}
\vspace{0.5cm}
\caption{\em Experimentally determined 
pixel thresholds and resulting
energy thresholds for different trigger modes.
The pixel thresholds are chosen such that the
rate of random 2-pixel coincidences is small 
(O(10\%)) compared
to the rate of triggers caused by air showers.
The energy thresholds are derived using Monte
Carlo simulations; the thresholds refer to
$\gamma$-rays and are defined as the energy
with the peak differential counting rate for
a source with a spectral index similar to the
cosmic-ray spectrum, for vertical incidence.}
\label{tab_thresh}
\end{center}
\end{table}
Table~\ref{tab_thresh} summarizes the typical
pixel trigger thresholds achieved for the 
different modes of operation. 
To relate these pixel thresholds to energy
thresholds, a Monte Carlo simulation of the 
system response is required.
Since the detailed
Monte Carlo simulation of the system is still under
development, a
fast simulation tool using parametrized trigger efficiencies
was used \cite{fast_sim}. Fig.~\ref{fig_drates} shows the
resulting differential detection rates for four cases:
(1) the 5-telescope IACT array, 
(2) the present 4-telescope
array with CT3 required in the analysis, in either case
with a 2-telescope trigger and a pixel threshold of 8 photoelectrons,
(3) the single telescope CT3 with a next-neighbor trigger
and a pixel threshold of 15 photoelectrons, and (4) CT3
with a 22 photoelectron threshold. While the fast simulation
does not reach the accuracy of a full Monte Carlo simulation,
it is expected to reproduce the relative performance rather
well. 
One notices that at large energies the effective area and hence
the differential detection rate of the system is not much larger
than that of a single telescope; the gain in total detection rate
results mainly from the lower threshold.
The resulting energy thresholds are included in 
Table~\ref{tab_thresh}; they should be reliable within 20\%. 
The threshold for the 
IACT system is about 1/2 of that of an 
equivalent (without next-neighbor trigger) single telescope.
The 8 photoelectron threshold, which was used for all system
data discussed here, results in a global trigger rate 
of 12~Hz.

Many other aspects of the multi-telescope trigger
coincidence have been investigated, such as
the dependence of rates on zenith angle or 
their dependence on the relative pointing
of the individual telescopes; more details 
will be published elsewhere. 

\section{Calibration and Data Analysis}

The analysis of data generated by the HEGRA IACT
system builds upon the well-known techniques 
for image analysis of single-telescope data,
augmented by new developments in certain areas,
such as an improved
geometry calibration of the telescopes,
the procedures to extract the 
amplitude information from the Flash-ADC data,
and the techniques for stereoscopic reconstruction.

\begin{figure}[tb]
\begin{center}
\mbox{
\epsfxsize14.0cm
\epsffile{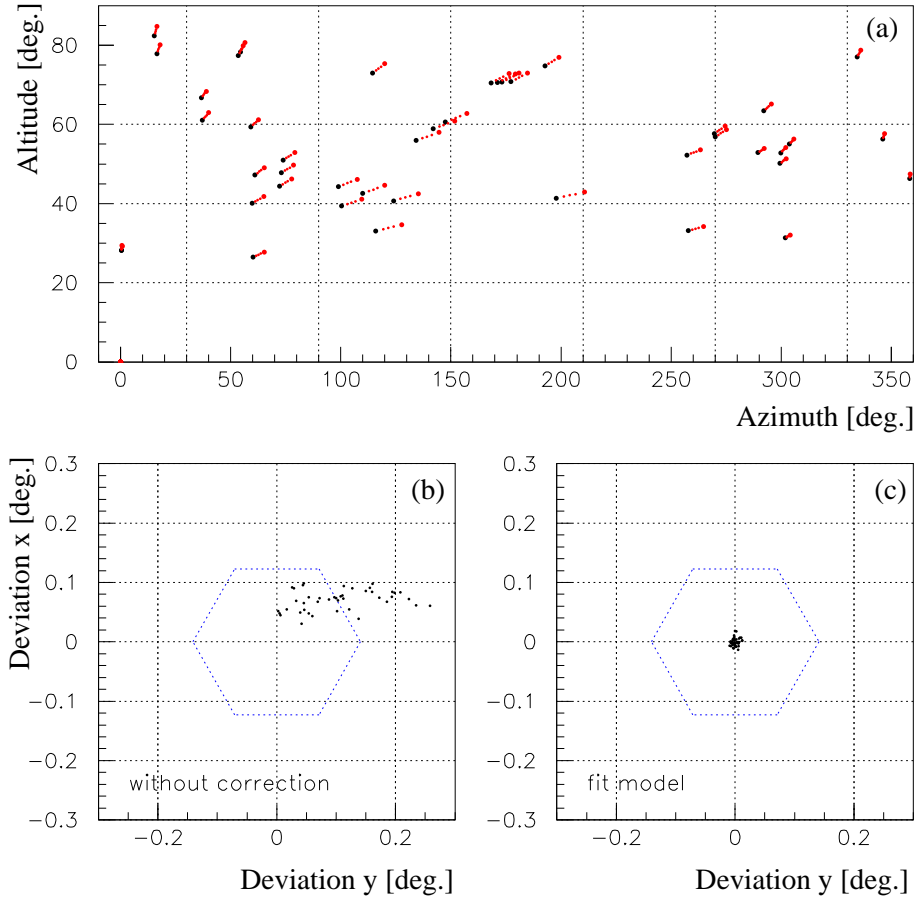}}
\end{center}
\caption
{\em (a) Pointing deviations of one of the HEGRA telescopes 
as a function of azimuth and altitude, determined using 
images of bright stars. The (enlarged) deviations
are indicated as vectors. (b) Measured deviations from
the nominal pointing for an ensemble of stars
(in the camera coordinate system). For comparison,
the outline of a 0.25$^\circ$ camera pixel is superimposed.
(c) Deviations after correction using the model discussed in the 
text.}
\label{fig_pointing}
\end{figure}

The stereoscopic reconstruction of air showers,
with its resolution in the mrad-range, is very
sensitive to deviations in the pointing of the
telescopes. In the past, pointing deviations of
up to one pixel size were observed
for some of the HEGRA telescopes, which is
absolutely unacceptable for a proper 
stereoscopic reconstruction.
Alignment procedures developed earlier within
HEGRA were extended and refined to cope with this
problem. A first step after the installation of 
a telescope is the alignment of the vertical axis
and a survey of the telescope by means of a 
theodolite. The final alignment is then based on
so-called ``point runs''. In these runs, the
image of a star is scanned in fine lines 
across one or more pixels, and the DC pixel
currents are measured. From those scans, both the
center of gravity of the image - and hence any
pointing deviations - and the point spread function
can be determined. To provide a complete map of
pointing deviations, the procedure is repeated
with many stars distributed over the entire
sky. Figs.~\ref{fig_pointing}(a),(b) illustrate
the differences between the nominal
positions of the stars and their actual images 
measured using the pixel currents, 
as a function of altitude and azimuth.
These data are then fit to a model of 
pointing deviations including atmospheric
refraction and, as free parameters, the 
bending of the camera masts (proportional
to $\cos(altitude)$), an offset of the camera
from the optical axis, a deviation of the azimuth
axis from vertical alignment, offsets in the
shaft encoder values, and 2nd harmonics for 
the shaft encoder response, caused e.g. by
eccentricity of the axes or gears. This model
provides a consistent description of the 
pointing data; after corrections, all stars
appear at their nominal positions, with an 
rms deviation of less than $0.005^\circ$
(Fig.~\ref{fig_pointing}(c)). Exact pointing
data exist so far only for 3 of the 4 telescopes
used here.

While the pointing calibration is repeated 
rather infrequently - typically a few times
per year - calibration data relevant for
camera and electronics operation are collected
regularly during data taking. 
In  particular, runs with a scintillator
light source pulsed by a laser are used to 
flat-field the camera, and to calibrate pixel timing.

\begin{figure}[tb]
\begin{center}
\mbox{
\epsfysize8cm
\epsffile{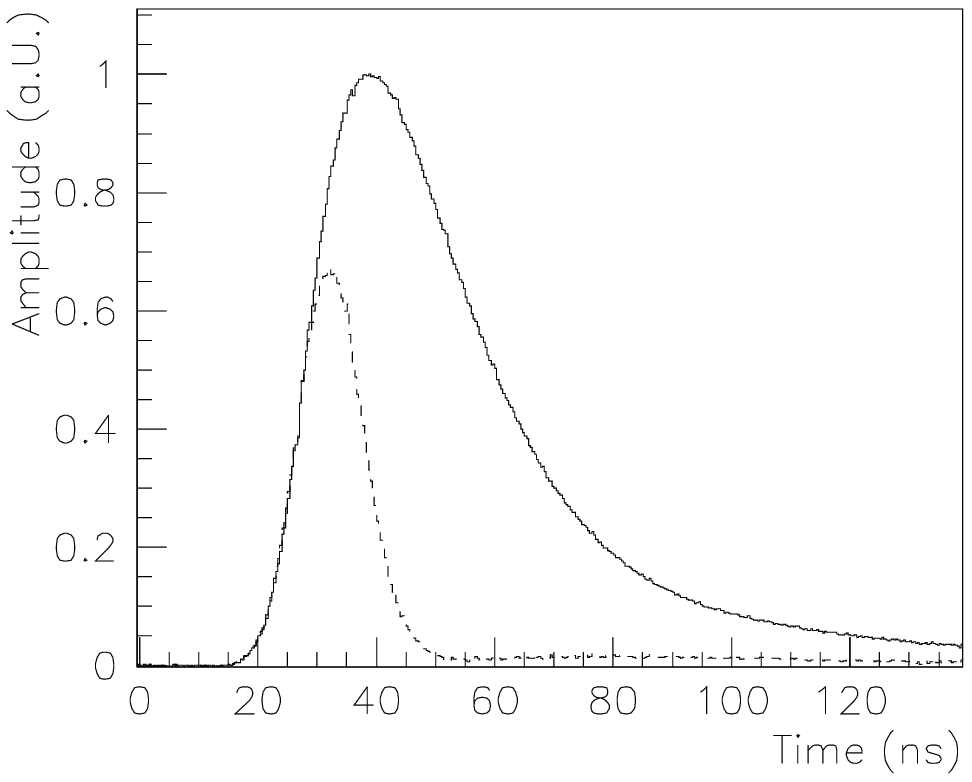}}
\end{center}
\caption
{\em PMT signal before and after
digital deconvolution of the Flash-ADC data,
averaged over many pulses.}
\label{fig_fadc}
\end{figure}

For the analysis for the Flash-ADC data, the
following procedure emerged: pedestal values are
continuously tracked and updated based on Flash-ADC samples
before and after the shower signal. For signals
which do not saturate the Flash-ADC dynamic range,
the signal is digitally deconvoluted. The 
deconvolution reverses the effect of signal
shapers in front of the Flash-ADC,
which are required to suppress signal components above the
Nyquist frequency. The deconvolution
shortens the signal and hence the effective 
integration time (Fig.~\ref{fig_fadc}), thereby
reducing the influence of sky noise. The deconvolution
assumes a linear response and cannot be applied
to signals saturating the dynamic range of the
8-bit Flash-ADC. For overflow pulses, the area under
the truncated pulse (i.e., the sum of the Flash-ADC
samples) is still a monotonic function
of the amplitude. A look-up table is used to
convert the area into an amplitude, extending
the linear dynamic range by a factor of more than 2, 
up to about 500 photoelectrons per pixel.

A conversion factor relating the ADC signal 
to the number of photoelectrons has been derived
with two independent techniques, once using the
width of the laser calibration signal (which is
essentially governed by photoelectron statistics), and once
using a pulsed light source shining a defined
amount of light onto the mirror. Both techniques
are in good agreement and give about 1 ADC count
per photoelectron.

Images are finally analyzed in terms of the usual
2nd-moment parameters, providing the center of gravity of
the images, their orientation, and their width
and length. A two-level tail cut is used to
selected pixels belonging to the image.
To be included, a pixel has either to show a signal equivalent
to 6 photoelectrons, or to have at least 3 photoelectrons
and a neighbor pixel with at least 6 photoelectrons.
Images are only accepted if they contain at least one
additional pixel adjacent to one of the trigger pixels.
 
\begin{figure}[tb]
\begin{center}
\mbox{
\epsfysize6cm
\epsffile{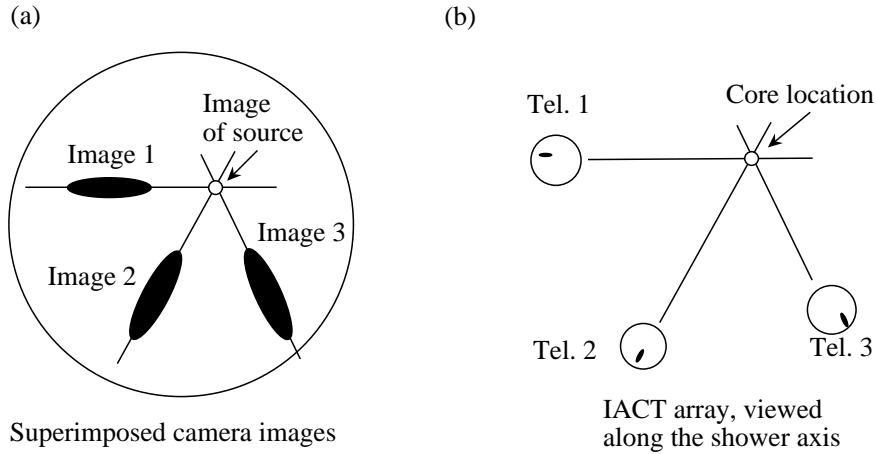}
}
\end{center}
\caption
{\em Reconstruction of (a) the shower direction and of (b) the 
core location from the images observed in the cameras.}
\label{fig_geom}
\end{figure}

The shower axis and core location are then derived
geometrically
(see, e.g., \cite{kohnle_paper,ulrichphd,ulrich_padua}, and
\cite{whipple_source}). 
Both the image of the source and the
point where the shower axis intersects the plane
of the mirror dish
have to lie on the symmetry axis of the image. The 
shower direction is hence derived by superimposing the 
images and intersecting their axes
(Fig.~\ref{fig_geom}(a)). The core location
is obtained by intersecting the image axes emerging from the
telescope locations (Fig.~\ref{fig_geom})(b)
(assuming the mirror planes of
all telescopes coincide; the extension to the general
case is straight forward). At present, the 
intersection points are calculated for each pair
of telescopes, and then averaged over all pairs,
weighted according to the angle between the views.
Advanced procedures and fits, which properly track
the errors of the image parameters, 
have been developed, but have not yet been
applied to the data.

First results are also available which use the
timing information in the Flash-ADC data for the
stereoscopic reconstruction, in addition to
the amplitude information; this topic is, however,
beyond the scope of this brief report.

\section{Performance for Crab Observations}

The analysis presented in the following is based on
about 11.7 (on-source) hours of observations of the Crab Nebula
with the four IACTs CT3,4,5,6. 
Zenith angles ranged from $5^\circ$ to $40^\circ$,
with a mean value of $20.2^\circ$.
About half of these
data were taken in ON-OFF mode, with equal shares
of time on-source and off-source. For the other half, the Crab
Nebula was displaced by $0.5^\circ$ from the optical
axis, and a region displaced symmetrically by the same amount
in the opposite direction is used to estimate backgrounds.

\begin{figure}[tb]
\begin{center}
\mbox{
\epsfysize10cm
\epsffile{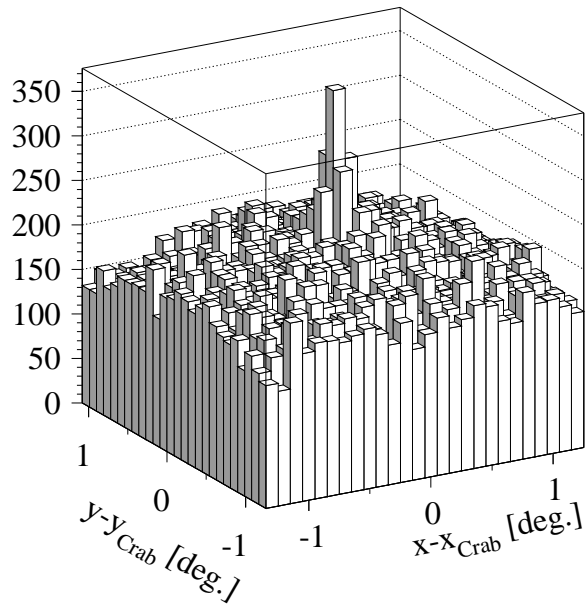}
}
\end{center}
\caption
{\em Distribution of the reconstructed shower
directions relative to the direction to the
Crab Nebula, for events where at least
two telescopes triggered, for 11.7 hours of
on-source observations, before cuts.}
\label{fig_xy}
\end{figure}

\begin{figure}[tb]
\begin{center}
\mbox{
\epsfysize11cm
\epsffile{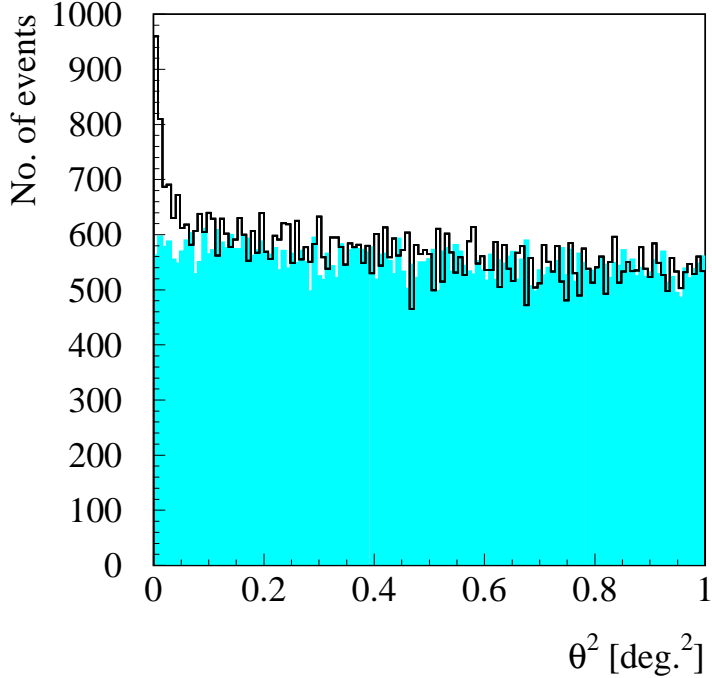}
}
\end{center}
\caption
{\em Line: distribution $dN/d\theta^2$ of events in the
square of the angle $\theta$ relative to the direction  to
the Crab Nebula. The  shaded histogram shows
the background derived from off-source runs and 
analyses with a shifted source position (see text).}
\label{fig_theta}
\end{figure}

Fig.~\ref{fig_xy} shows the distribution in the direction
of reconstructed shower axes, without any additional
cuts. An excess from the direction of the Crab Nebula
is already visible in this raw data. 
For a quantitative
analysis, we plot the distribution in the angle $\theta$
between
the shower axis and the direction to the Crab Nebula; shown in 
Fig.~\ref{fig_theta} is $dN/d\theta^2$. Since $\theta$
is an angle in space, one expects for the uniform
background from charged cosmic rays a flat distribution,
\begin{equation}
\left( {dN \over d\theta^2} \right)_{CR}
= const.~~~,
\end{equation}
whereas a point source at the origin generates an excess
\begin{equation}
\left( {dN \over d\theta^2} \right)_{Source}
\sim e^{-\theta^2/2 \sigma^2}~~~~,
\end{equation}
with the (projected) angular resolution $\sigma$.
Fig.~\ref{fig_theta} illustrates that the distribution
of events is indeed flat over a wide range in
$\theta^2$, with the signal confined to a narrow
spike at $\theta \approx 0$.

It is instructive to compare these (uncut) data with typical data reported 
for conventional single telescopes - the 
equivalent plot is the distribution in the angle $\alpha$
between the image axis and the direction to the camera
center, i.e., the source (see Fig.~\ref{ct3_crab}).
In terms of the signal to noise ratio, the uncut system data
are not too far from the single-telescope data where tight
image cuts have been applied; in the raw single-telescope
data the Crab signal is barely visible.

The advantages of the stereoscopic technique are related to the fact
the the signal is confined to a small region (about 
1\%) of the available two-dimensional phase space, whereas in the typical
$\alpha$-distribution the signals extend over about
10\% of the $\alpha$-range. In particular,
\begin{itemize}
\item due to the concentration of the signal events,
a highly significant  peak is seen already 
in the 11.7~h of raw system
data
\item due to the flat distribution of background 
 in two dimensions, 
 the background under 
the signal can be estimated reliably
even without dedicated off-source runs.
\end{itemize}

\begin{figure}[tb]
\begin{center}
\mbox{
\epsfysize10cm
\epsffile{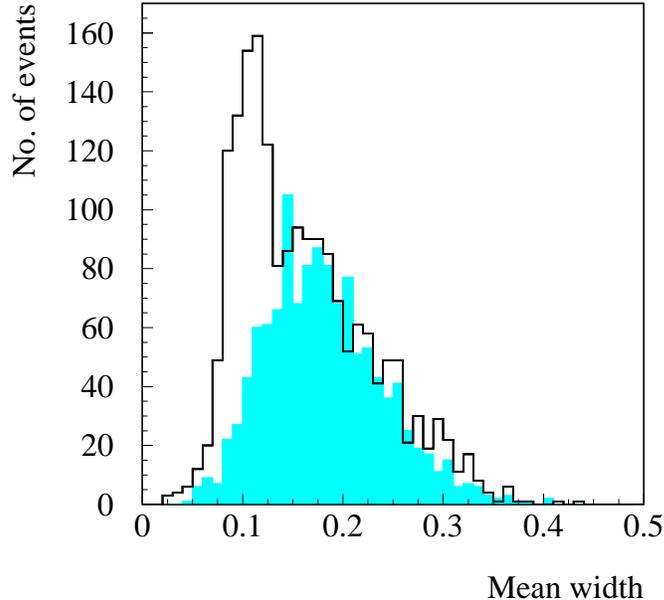}
}
\end{center}
\caption
{\em 
Line: distribution of the 
mean width $\bar w$, averaged
over all telescopes which triggered in a given event,
for showers from the direction of
the Crab Nebula. The  shaded histogram shows
the background derived from off-source runs and 
analyses with a shifted source position.}
\label{fig_meanw}
\end{figure}

Apart from the information concerning the shower
direction - for system data, contained directly in
the direction reconstructed event-by-event, for 
single-telescope data contained e.g. in the variable
$\alpha$ and the distance $d$ between the image and
the camera center - image shapes are used to reject
hadronic showers. Relevant image parameters are
the width $w$ of the images, their length $l$,
the degree of concentration of the image, etc.
Figure~\ref{fig_meanw} shows, e.g., the 
distribution in the mean width $\bar w$
- averaged over all telescopes which triggered in a given
event - for showers with directions within 0.13$^\circ$
from the Crab. As expected for $\gamma$-ray showers, the
excess events have small values of $\bar w$.

Given that the different image parameters are strongly 
correlated, and dependent on the amplitude of an
image, the optimization of cuts is non-trivial
even in the case of a single telescope. The
number of parameters increases with the number
of telescopes, rendering the problem even more
complex. A simple cut e.g. on the widths of all
images is a rather inefficient procedure, since 
some images - those with large amplitudes and
optimal distances around 100~m from the shower
core - provide a large separation power, whereas
faint images differ little between hadronic and
photon-induced showers and a cut merely reduces
the efficiency for photons, without a corresponding
gain in signal-to-noise. Therefore, a different
procedure was followed: Monte-Carlo events were
used to parametrize the distribution in width $w$
and length $l$ as a function of the nature of the
incident particle, of the amplitude $A$ of the image,
of the distance $D$ to the shower core, and of the
zenith angle $z$ under which the shower was observed.
Lacking sufficient Monte Carlo statistics, the
joint distribution was factored into a $w$-dependent
and a $l$-dependent term, neglecting their correlation:
\begin{equation}
\rho(w,l|A,D,z) = \rho_w(w|A,D,z) \rho_l(l|A,D,z)~~~.
\end{equation}
A `probability' $p$ for a given hypothesis - $\gamma$-ray
or cosmic-ray shower - was calculated by multiplying
the terms for the individual telescope images $i$,
with their image parameters $w_i$ and $l_i$ and the 
image amplitude $A_i$
\begin{equation}
p = \prod_i \rho_w(w_i|A_i,D_i,z) \rho_l(l_i|A_i,D_i,z)~~~~.
\end{equation}
The distance $D_i$ between telescope $i$ and the shower
core is calculated using the stereoscopic reconstruction of the
core location. 
Events were then selected by requiring that the `probability'
for the $\gamma$-ray hypothesis exceeds the `probability'
for the cosmic-ray hypothesis by a certain factor.
This procedure avoids the potentially rather inefficient
cuts on individual image parameters. 

Of course, rather
than using Monte-Carlo cosmic-ray events, one can use
real events from off-source runs; both choices gave
similar performance.

\begin{figure}[tb]
\begin{center}
\mbox{
\epsfysize11cm
\epsffile{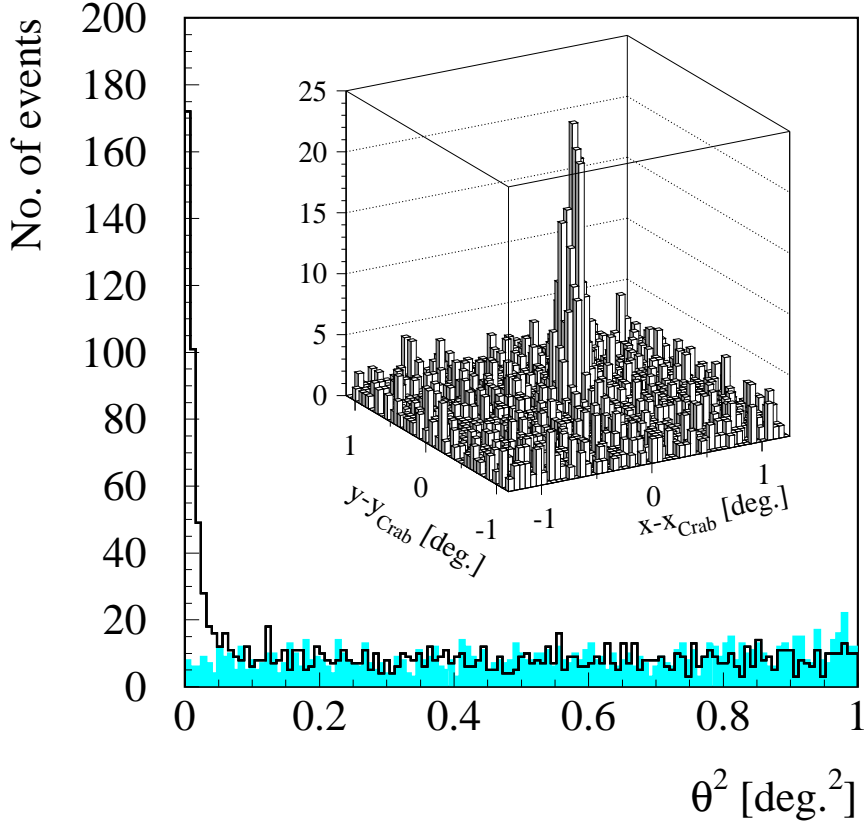}
}
\end{center}
\caption
{\em Line: distribution of reconstructed shower directions,
relative to the direction to the Crab Nebula, after
cuts on the event shapes. The shaded histogram shows
the background derived from off-source runs and 
analyses with a shifted source position.
The small insert shows the two-dimensional distribution
of shower directions.}
\label{fig_cut}
\end{figure}

\begin{figure}[tb]
\begin{center}
\mbox{
\epsfysize11cm
\epsffile{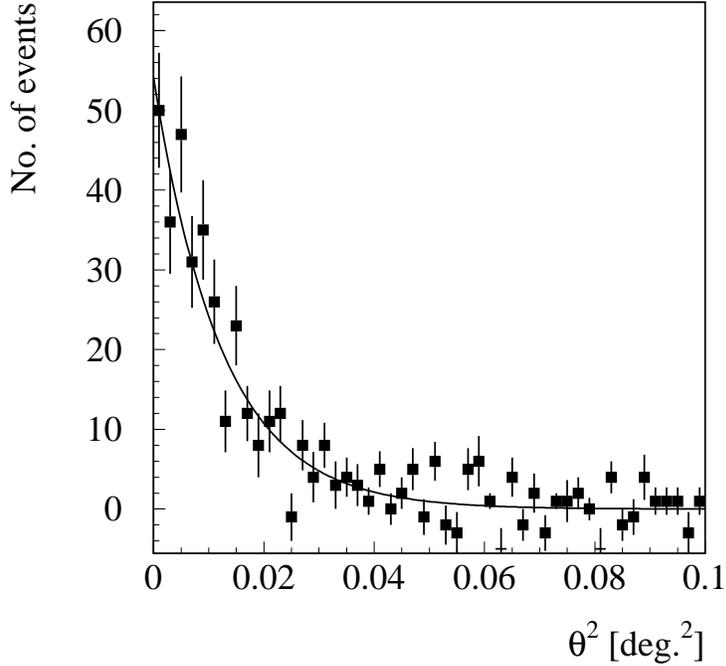}
}
\end{center}
\caption
{\em 
Expanded view of the signal region,
showing the background-subtracted peak with
an exponential fit corresponding to a projected
angular resolution of $0.08^\circ$.}
\label{fig_diff}
\end{figure}

Figs.~\ref{fig_cut},\ref{fig_diff} show the directional distribution
of events after this cut, with the additional requirement
that at least two telescopes have images with 50 or more
photoelectrons. The projected rms angular resolution derived
from the peak is $0.08^\circ$
(Fig.~\ref{fig_diff}); for optimum signal-to-noise,
one should hence select a region with a radius of about
$0.13^\circ$ around the source. Table~\ref{tab_rates}
lists the resulting rates and efficiencies. The shape
cuts discussed above allow to reduce comic-ray background
by almost a factor 100, while maintaining a 40\% efficiency
for $\gamma$-rays. With simpler cutting procedures, based
e.g. on the mean width $\bar w$ of all telescope images, 
and on their
mean length and concentration, 
background rejection is about a factor 2 worse, but still
sizeable.

We note that for short observation times of one hour, 
one expects about one background
event in the signal region, 
whereas a source with a flux of the Crab generates
about 20 events. This features allows e.g. sky surveys with
an observation time of about 1~h per $2^\circ$ by $2^\circ$
region (the effective field of view of the system), at a 
sensitivity of about 30\% of the Crab flux. For longer-term
observations - 100 h on-source corresponding to one or
two months of data taking - a source of 3\% of the Crab flux should 
still generate a 5-sigma signal, on top of an average
background of about 120 events \footnote{Since the background
is flat over a region which is very large compared
to the signal region, the expected background can be 
estimated with good statistical precision. This is
in contrast to single-telescope observations, where
the background is usually determined from an off-source sample,
and has similar statistical uncertainty as the signal sample.}.
For comparison, the single telescope CT3 provides
a detection limit of 15\% to 20\% of the Crab flux after
100~h on-source
(see Fig.~\ref{ct3_crab}). 
A uniform extended source of diameter $1^\circ$
is detectable by the system if the total flux exceeds
15\% of the Crab flux, assuming 100~h of on-source data.
In this case, an equivalent amount of off-source data is
required for background subtraction.
\begin{table}
\begin{center}
\begin{tabular}{|l|c|c|c|}
\hline
 & Before shape cuts & After shape cuts & Selection eff.\\
\hline \hline
Signal & 55/h & 23/h & 0.42\\
(MC) & (58/h) & (29/h) & (0.5) \\
\hline
Background & 105/h & 1.2/h & 0.011 \\
(MC) & (95/h) & (3/h) & (0.03) \\
\hline
\end{tabular}
\vspace{0.5cm}
\caption{\em Detection rates for Crab observations
with the 4-telescope system, before and after shape
cuts, and efficiency of the shape cuts. Rates 
given refer to a circle with radius $0.13^\circ$
around the source location. The numbers
in parentheses represent Monte Carlo estimates
(see text); for the $\gamma$-ray source a
flux of $10^{-11}$/cm$^2$s above 1 TeV is assumed,
and a differential spectral index of 2.7.
For the background, only the cosmic-ray proton
component is considered; heavier nuclei have
significantly reduced trigger probabilities.}
\label{tab_rates}
\end{center}
\end{table}

To provide a consistency check,
Table~\ref{tab_rates} also includes Monte Carlo based
estimates for the detection rates. For the $\gamma$-ray
source a differential spectral index of 2.7 is assumed,
and a flux of $10^{-11}$/cm$^2$s above 1 TeV.
Recent single-telescope HEGRA measurements
give values for the Crab flux above 1~TeV of $(0.8 \pm 0.3) 
\cdot 10^{-11}$/cm$^2$
\cite{crab5} and $1.5^{+1.0}_{-0.5} \cdot 10^{-11}$/cm$^2$ \cite{crab6},
with a spectral index of $2.7 \pm 0.3$, and
$0.77^{+0.47}_{-0.13} \cdot 10^{-11}$/cm$^2$ \cite{crab7}
above 1.5~TeV.
The raw detection rates given in the table are based on the
fast simulation tool using parametrized trigger efficiencies. 
The Monte Carlo
estimates for the rates before cuts should be reliable within 50\%. 
Given these
caveats, the agreement of simulated and measured rates
before cuts is surprisingly good. 

The Monte Carlo selection efficiencies quoted 
in the table refer to early simulations,
which were carried out using a slightly different configuration and
different selection criteria, and should only serve to illustrate
approximate magnitudes; the cuts used in the present data
analysis are obviously somewhat tighter.
The angular resolution has been
simulated~\cite{ulrich_padua} using the identical algorithm used for 
shower reconstruction, and the predicted value of 0.1$^\circ$
is in good agreement with the measured (projected)
resolution of $0.08^\circ$ after cuts, or $0.09^\circ$
before cuts. (The cuts bias towards well-measured narrow
showers, resulting in a slightly improved resolution.)

\section{Conclusion}

In summary, it seems appropriate to state that while
both the hardware of the HEGRA IACT array and the
analysis techniques are still evolving and further
improvements are to be expected, existing data clearly
demonstrate the power and the potential of the 
stereoscopic approach. The detection rates and
angular resolutions are consistent with expectations based on
Monte-Carlo studies. Data clearly demonstrate
the lower trigger thresholds achievable with a 
telescope coincidence trigger, the high level of
suppression of the cosmic-ray background, and the 
superior angular resolution.

\section*{Acknowledgements}

The support of the German Ministry for Research 
and Technology BMBF and of the Spanish Research Council
CYCIT is gratefully acknowledged. We thank the Instituto
de Astrofisica de Canarias for the use of the site and
for providing excellent working conditions. We gratefully
acknowledge the technical support staff of Heidelberg,
Kiel, Munich, and Yerevan.

\end{document}